# Superconducting detector for visible and near-infrared quantum emitters

**V. Vorobyov,**[1,2,3,*] **A. Kazakov,**[4] **V. Soshenko,**[1,2,5]
**A. Korneev,**[1,3,4] **M. Y. Shalaginov,**[7] **S. Bolshedvorskii,**[1,3,5]
**V.N. Sorokin,**[5] **A. Divochiy,**[8] **Yu. Vakhtomin,**[4,8]
**K. V. Smirnov,**[4,8] **B. Voronov,**[4] **V. M. Shalaev,**[7] **A. Akimov,**[1,5,6]
AND **G. Goltsman**[4,9]

[1]*Russian Quantum Center, 143025 Moscow Region, Russia*
[2]*Photonic Nano-Meta Technologies LLC, 141009 Moscow Region, Russia*
[3]*Moscow Institute of Physics and Technology, 141700 Moscow Region, Russia*
[4]*Moscow State Pedagogical University, 119991 Moscow, Russia*
[5]*P.N. Lebedev Physical Institute, 119991 Moscow, Russia*
[6]*Texas A & M University, College Station, TX 77843, USA*
[7]*School of Electrical & Computer Engineering, Birck Nanotechnology Center, and Purdue Quantum Center, Purdue University, West Lafayette, IN 47907, USA*
[8]*Scontel CJSC, 119021 Moscow, Russia*
[9]*Higher School of Economics (National Research University), 101000 Moscow, Russia*
[*]*mrvorobus@gmail.com*

**Abstract:** Further development of quantum emitter based communication and sensing applications intrinsically depends on the availability of robust single-photon detectors. Here, we demonstrate a new generation of superconducting single-photon detectors specifically optimized for the 500-1100 nm wavelength range, which overlaps with the emission spectrum of many interesting solid-state atom-like systems, such as nitrogen-vacancy and silicon-vacancy centers in diamond. The fabricated detectors have a wide dynamic range (up to 350 million counts per second), low dark count rate (down to 0.1 counts per second), excellent jitter (62 ps), and the possibility of on-chip integration with a quantum emitter. In addition to performance characterization, we tested the detectors in real experimental conditions involving nanodiamond nitrogen-vacancy emitters enhanced by a hyperbolic metamaterial.

## 1. Introduction

Recent advances in nanophotonics and plasmonics have enabled new avenues for the development of quantum integrated circuitry based on solid–state spin systems [1–7]. A good example of a well–developed solid–state systems is a nitrogen–vacancy (NV) color center in diamond [8]. Recent achievements with NVs include the demonstration of a quantum memory bit with a lifetime exceeding one second [9], quantum registers [10], and the entanglement of two independent centers [11]. In addition, available diamond manufacturing techniques enable the fabrication of photonic structures directly out of single–crystal diamond substrates [1, 12]. The development of elements for integrated quantum nanophotonic circuitry based on NV or alternatively silicon–vacancy (SiV) centers [13–15] is promising for all–optical [16, 17] and quantum information processing [18, 19]. Moreover, NV color centers have been widely used in various nanoscale sensing applications. These applications include biocompatible thermometry [20], electric and magnetic fields sensing [21, 22], high–resolution magnetic scanning microscopy, and magnetic imaging systems [23]. In these applications, NV centers provide electron–spin states which have long coherence times [24] that can be both initialized and read out optically. Gaining control over the single–spin system places new demands on detection devices, such as low detector dark counts and wide dynamic range. Currently, a key direction is to develop quantum emitters with high count rates by coupling the emitters to various types of emission enhancing structures, such as photonic–crystal cavities [25], ring resonators [26], plasmonic waveguides/cavities [2, 27], metamaterials [28] and metasurfaces [29], and the use of multiple color centers [1, 20, 30–33]. Superconducting single–photon detectors (SSPD) [34, 35] can offer a much wider dynamic range compared to available Si single–photon avalanche diodes (SPAD). The improvement in dynamic range comes from the much lower dead time along with an extremely low dark count rate. In addition, SSPDs offer cutting–edge timing resolution, which is comparable to or slightly better than the resolution of the best available avalanche photodetectors (APD) [36].

Previously, SSPDs have shown outstanding performance at telecom frequencies (i.e. 1550 nm) [37], critical for applications such as quantum cryptography [38, 39] and for a number of applications in photonics [40–42]. Moreover, SSPDs could be integrated onto a photonic chip [43–46]. In this work we have developed and characterized an SSPD working in the visible and near–IR spectral ranges (500–1100 nm). Efficient detectors operating in this spectral range are of significant importance for measuring numerous quantum emitters including NV centers. Therefore, we tested the detectors in a real experiment involving nanodiamond nitrogen-vacancy centers coupled to a hyperbolic metamaterial.

## 2. Results

### 2.1. SSPD design

The demands of quantum optics for color centers requires reconsidering and adapting the SSPD architecture. Based on the ideas suggested earlier [47, 48] we have developed a new generation of SSPD detectors operating in the spectral range of 500 – 1100 nm. The SSPD was composed of a 4–nm–thick NbN film on a Si substrate with a 160 nm layer of $SiO_2$. This layer serves as a weak broadband cavity, which is introduced to improve photon absorption at the operational wavelengths (Fig. 1(a)). The active element of the detector is 120–nm–wide NbN stripe in the form of a meander, which covers a $7 \times 7 \, \mu m^2$ area with a filling factor of 0.6. The filling factor is defined as a ratio between the area covered with the meander and the total device area. Each detector is coupled with single–mode optical fiber (SM–800, Thorlabs), which has a mode field diameter of 5.6 μm at an 830 nm wavelength.

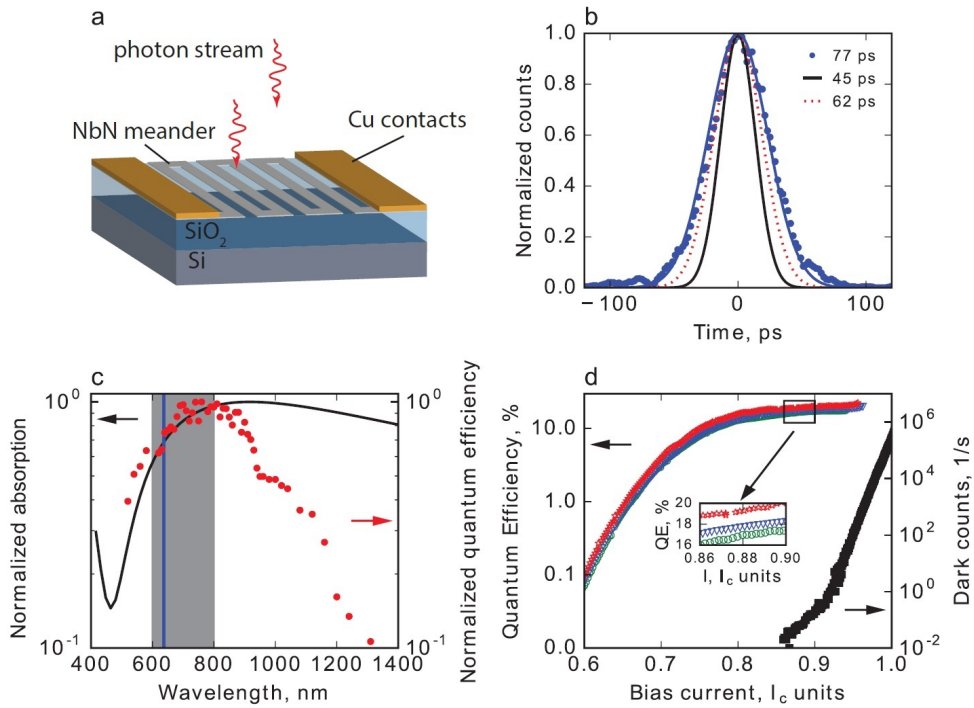

Fig. 1. a) Schematic of SSPD consisting of Cu contacts, 4–nm–thick NbN meander, and 160 nm thick $SiO_2$, forming a cavity on top of 0.5–mm–thick Si substrate. b) SSPD jitter (doted red line) was found to be 62 ps. Jitter has been calculated as a square root of the difference between squared FWHM of the counts histogram (blue curve) and squared nominal laser pulse duration (black line). c) Red dots show spectral sensitivity of SSPD for unpolarized light in the 500–1300 nm range at operating temperature of 4.2 K, measured for the photon flux of $10^8$ photons per second. Black curve shows calculated absorption spectrum of the SSPD with a 160 nm $SiO_2$ cavity. Gray area on the graph shows NV centers emission range. Blue line shows NV center zero–phonon line. d) Quantum efficiency at 633 nm wavelength (colored points) and dark count rate (black points) of SSPD versus detector bias current at operating temperature of 4.2 K. Red, blue and green curves in the inset stand for the cases of optimal polarization giving maximum count rate, unpolarized light and polarization giving minimum count rate, respectively. Bias currents are given in fractions of a critical current $I_c$, which is measured to be 29 μA.

## 2.2. SSPD performance characterization: jitter, quantum efficiency, dark count rate

One of the key features of our detector is its cutting–edge jitter performance. The jitter was measured using a pulsed laser with a pulse duration below 45 ps, and repetition rate of 200 MHz. We attenuated the laser to produce less than one photon per pulse. Fig 1(b) shows the histogram of the rising–edge time distribution measured using a digital oscilloscope (Tektronix DPO 70404C). The jitter, calculated as FWHM of the deconvolution of the measured histogram and a pulse shape, was found to be 62 ps.

Quantum efficiency (QE), i.e. the probability of detecting an incident photon, is presented in figures 1(c),(d). The measured QE (red points on Fig. 1(c)) have a broad distribution with a maximum at 700 nm. This distribution covers both the NV and SiV emission spectra as well as Rb and Cs resonance lines. The decrease in QE at shorter wavelengths is caused by the cavity which the SSPD is integrated with. While the decrease of the QE at wavelengths above 900 nm is caused by the reduction of the SSPD internal detection efficiency, i.e. a photon is absorbed but does not create a resistive state. Due to the device–construction the detector sensitivity depends on the incident light polarization [49]. The QE of our detector reaches 20% for optimal polarization, 18% for unpolarized light at the NV zero–phonon line (around 637 nm) and slightly improves toward 800 nm reaching a maximum value of 30%. Fig. 1(d) presents the QE (color points) for different light polarizations and dark count rates (black points) as a function of the detector bias current given in the units of detector critical current $I_c$. We define $I_c$ as the current at which the superconductivity in the device is permanently broken. For our detector operating at 4.2 K $I_c$ is 29 µA. From Fig. 1(d) we also see that our detector's quantum efficiency saturates at dark count rates as low as $10^{-1} s^{-1}$. This is possible due to the filtering–effect of the cooled single-mode optical fibers [50]. Bends in the fiber at 4.2 K introduce high losses of long–wavelength thermal photons from the room–temperature background. This approach enabled reduced dark counts across the entire detector operating range, 700 nm, in contrast to the 20 nm bandpass filter approach reported elsewhere [51].

### 2.3. SSPD test in a real experiment: collecting emission from an NV center coupled to a metamaterial

One example of strong modification of the NV center emission is observed when the emitters are coupled to hyperbolic metamaterials [28]. We used samples prepared in the manner described in Ref. [28] to compare the performance of our SSPD versus a Perkin–Elmer APD (SPCM–AQRH–14–FC). Nanodiamonds with a nominal size of 50 nm (Microdiamant AG) were spin–coated onto a hyperbolic metamaterial, which consisted of ten pairs of TiN/(Al,Sc)N layers, where each layer was 10 nm thick.

The experimental setup to study the NV centers luminescence is shown in Fig. 2(a). The diamond sample was pumped with a continuous–wave (Coherent, Compass 315M–100) or with a pulsed (PicoQuant, LDH–P–FA–530XL) 532 nm laser through a confocal microscope. The resulting fluorescence of the NV centers was collected into two independent channels which were coupled either to SSPDs or to APDs. Notch filters in combination with long–pass filters were used to eliminate the 532 nm pumping radiation and reduce the background signal of the substrate.

Using our detectors we measured the $g^{(2)}(\tau)$ autocorrelation function of a single NV center as well as the lifetime (see Fig. 2(b,c,d)).The $g^{(2)}(\tau)$ autocorrelation function statistics were collected during the 300 seconds of integration time for both the SSPD and APD detectors. To be able to do time–correlation measurements we used two detectors in combination with a beamsplitter. We used a two-detector scheme, since the lifetime of the NV center becomes comparable with the detector dead time due to the emission enhancement caused by the metamaterial.

We note that our SSPDs have a very low dark count rate, which in combination with the low jitter enables them to perform rapid and precise measurements of photon statistics with low signal–to–noise ratios as well as at high emission rates. Consequently, despite the lower efficiency of the SSPD, the $g^{(2)}(\tau)$ shows a larger drop near τ=0 than the response measured with APDs. To verify this conclusion, we performed modeling of the $g^{(2)}(\tau)$ function with realistic parameters. This model confirms, that low dark counts improve the quality of registered $g^{(2)}(\tau)$ even for relatively low quantum efficiencies (Fig. 2(e), see also supporting information). This improvement clearly demonstrates the advantage of our SSPD.

Lifetime measurements of a single NV center also clearly demonstrate the advantages of our SSPD (Fig 2d). In the figure, one can see the long tail which corresponds to the NV center's spontaneous decay and fast decay processes due to the residual luminescence of the HMM substrate, which is not seen with APD.

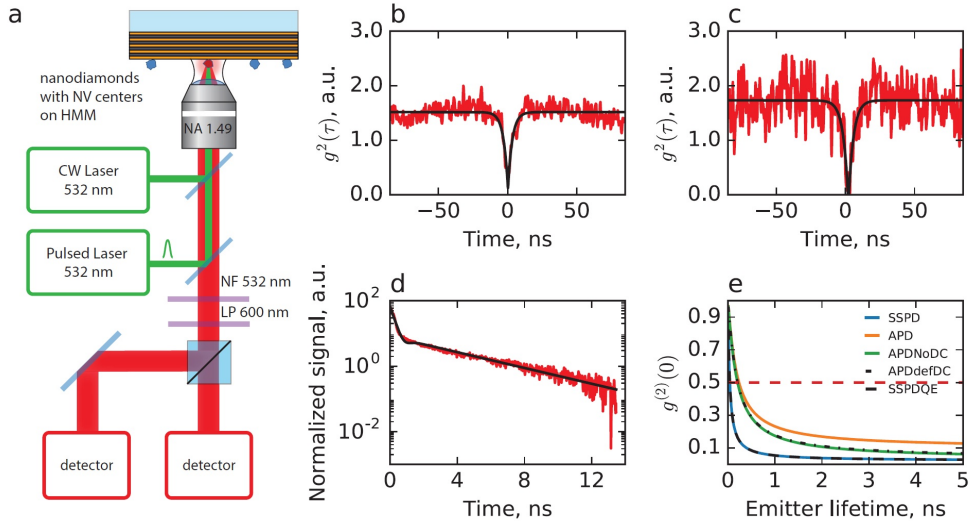

Fig. 2. a) Schematic of experimental setup. b),c) Comparison of $g^{(2)}(\tau)$ autocorrelation function obtained by a commercially available APD (b) and an SSPD (c). d) Photoluminescence decay for a single nanodiamond NV center on top of the hyperbolic metamaterial. e) Simulation of the dependence of the $g^{(2)}$ function at $\tau = 0$ on quantum emitter lifetime (see supporting information for more details). Orange line – conventional APD with QE = 0.6 and darkcounts level 1500 cps, jitter 300 ps; green line – idealized APD with dark counts 0.1 cps, black dash doted line – APD with dark counts specified in data sheet. Blue line – SSPD with QE = 0.2, jitter = 0.06 ns, black line – SSPD with high QE = 0.6. Signal–to–noise ratio was chosen to be 100 in order to take into account background compensation procedure (see Appendix).

### 2.4. Detector operation at high counting rates

Another important parameter, which characterizes the performance of a single–photon detector is the maximum possible count rate one can observe with the detector. Additionally, the counting behavior under strong illumination of laser light has recently become of high interest due to the realization of high power attacks on quantum cryptography lines [52, 53].

Fig. 3(a) presents the dependence of the maximum count rate on the bias current (blue line). The maximum achievable count rate strongly depends on the operational current of the detector and features a rapid increase at a certain bias current with subsequent decay. From the detector dead time of 3 ns, one could estimate the maximum count rate is 280 MHz while Fig. 3(a) shows maximum rate of 270 MHz. It is also interesting to note that the maximum count rate is achieved at a very low QE (red points in Fig 3(a)). However, for the SSPD due to its low dark counts the low QE does not lead to a pure signal–to–noise ratio, as is shown in supporting information Fig. 6.

The maximum count rate versus detector bias current (blue points in Fig. 3(a)) has a sharp rising edge with linear behavior before and after the step. One could assume that this step–like behavior corresponds to the detector switching to multi–photon counting mode. Nevertheless, the fact that at high photon rates the dependence has the same slope indicates that the detector is still operating in the single–photon regime.

Additionally, using our model we can confirm the single–photon character of the detector by comparing the measured probability to detect the next photon following a successful detection (see Fig. 3(d)). The non–linear behavior of the detector counts was qualitatively explained in a model by Kerman [54]. Here, we provide a probabilistic model of the feedback, giving further insight into the detector's operation at high counting rates. Following Kerman [54], we believe the step like behavior indicates a quick change in the detector quantum efficiency caused by a positive feedback reaction in the current stabilization circuit. When the detector absorbs a photon, a momentary (< 1 ns) circuit break appears followed by an exponential rise of the detector current to operating value. If the circuit breaks are not frequent and there is enough time between pulses for detector to stabilize, the mean current observed is close to operating value. For higher frequencies the supply circuit will tend to stabilize and average the current value between the on and off states, thereby modifying the operational current of the detector and its quantum efficiency. In our model we included this feedback along with realistic detector efficiency and detector current pulse decay time. The results of the modeling are represented in Fig. 3(b). The figure illustrates a comparison of our experimental data with our positive feedback model near the step–like rise in the count rate (supply current $0.62\,I_c$). One can see good agreement between the model prediction and the experimental results.

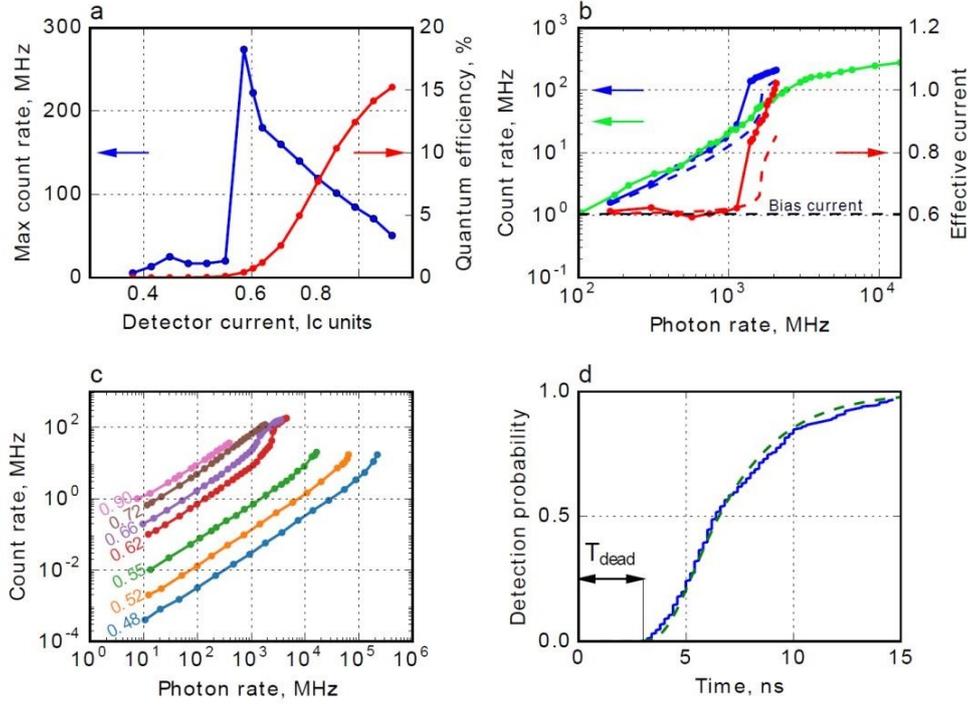

Fig. 3. a) Maximum count rate (blue) and quantum efficiency (red) of the detector at 532–nm wavelength as a function of detector bias current. b) Detector count rate (blue) and effective current (red): experimental data (solid lines) and modeling results (dashed lines); bias current is $0.62\,I_c$. Green line is for constant voltage regime corresponding to bias current of $0.62\,I_c$ set at 1 MHz count rate. Effective current was measured by exponential fitting of the detector signal decay. c) Detector count rate as a function of input photon rate at 532 nm for different bias current(shown on the plot as a fraction of critical current) d) Next photon detection probability under exposure (bias current is $0.62\,I_c$, count rate is 123 MHz); dead time is 3 ns; after 10 ns detector is almost fully recovered.

It can be seen from Fig. 3(b) that when the count rate rises the detector's operating current also rises. This implies that the detector efficiency increases monotonically as a function of current, leading to a count rate and operating current increase. Such positive feed–back is limited by quantum efficiency saturation (Fig. 3(c)). Interestingly at lower operational currents our model predicts divergent behavior in the feedback which causes the detector to latch effectively leading to a lower maximum achievable counting rate. This prediction was confirmed experimentally (see Fig. 3(c)).

The singularity in the detector operating in the electrical current stabilization regime could be of interest for fiber based click sensors such as alarm and other monitoring applications. Under voltage stabilization our sensor shows a higher dynamic range (0–350 Mcps) with linear counting up to 200 Mcps.

In the voltage stabilization regime, the step–like behavior of the counts is not observed due to the absence of the positive feedback. Note that while in the voltage stabilization regime, saturation of the counts is still observed after $10^8$ counts per second, resulting in a linear regime which is much wider than observed in the current stabilization regime. The full dynamic range in this regime is limited to 350 MHz.

## 3. Conclusions

In summary, we have presented a new generation of SSPD detectors, suitable for photonics research in the visible and near–infrared region. We have shown a dynamic range up to 350 MHz in non–linear and 200 MHz in linear modes, jitter of 62 ps and dark counts as low as 0.1 count per second which makes this detector very attractive for researchers in the field of quantum optics and quantum information processing. We also carefully investigated the step–like behavior in the detector's quantum efficiency, providing a simple qualitative model for this behavior. To assess our detectors real-world performance we conducted an experiment involving nanodiamond nitrogen-vacancy emitters enhanced by a hyperbolic metamaterial. The experiment verified our detectors are indeed well adapted to systems of this nature. To further improve the performance of our detector we will embed the SSPD in an improved cavity as described elsewhere [48]. Moreover, we believe this work will open an avenue to create fully integrated scalable photonic chips which combine NV centers, photonic waveguides, and detectors with near 100% quantum efficiency [45] on a single chip.

Additionally, we believe that due to its exceptionally high dynamic range, our detector could be used beyond quantum applications. One such application could be space–based LIDAR [55], system where the detector should be able to detect weakly reflected laser pulses with single–photon precision if the signals have a wide dynamic range. Another example is fiber–based optical sensing of strain, temperature, and gas concentration. Current sensors rely on precision measurements of the transmittance differences of optical channels and suffer from detector noise [56, 57]. Hence they can substantially benefit from our SSPD design.

**Appendix**

**Measurement of SSPD quantum efficiency, jitter, and g$^{(2)}(\tau)$ function**

We define quantum efficiency $QE$ as the ratio of the photon count rate $CPS$ to the photon flux $N$ incident on the detector:

$$QE = \frac{CPS}{N}. \tag{1}$$

To determine the quantum efficiency near NV zero-phonon line we coupled the detector to 633 nm He-Ne laser. The intensity of the laser was measured using a power meter (NewFocus 2001, Thorlabs PM 100). The laser beam was attenuated to produce 31.4 pW of optical power corresponding to $10^8$ photons per second. Fig. 1(d) in the main text shows $QE$ and dark count rate of the detectors versus bias current in the fractions of critical current $I_c$. To get unpolarized light we used monochromator with incident lamp operating at the same wavelength as laser.

We calculated absorption of NbN meander with a 160 nm $SiO_2$ layer on Si substrate (black line in Fig. 1(c)) using 1D array transfer matrix method [58] following the procedure described in Ref. [59]. Wavelength dependencies of Si and $SiO_2$ refractive indeces were taken from Ref. [60]. To get experimental data of SSPDs spectral sensitivity (red dots in Fig. 1(c)) we used a monochromator with incident lamp. We calibrated its output to produce $10^8$ photons per second for each wavelength in the range from 500 nm to 1310 nm using NewFocus 2001 and NewFocus 2011 power meters and attenuators to measure ultra low optical power. Experimental QE goes down after 900 nm because of detector internal QE falling with photon energy and not each absorbed photon produces a pulse.

To measure SSPD jitter we attenuated laser (Intec) to produce less than one photon per pulse, connected sync output of our laser to trigger channel of digital oscilloscope (Tektronix DPO 70404C) and plotted a histogram of SSPD pulse-rising-edge time distribution.

In $g^{(2)}(\tau)$ function measurement we used two detectors one of which was connected to "start" channel of TCSPC module(PicoHarp) and another SSPD was connected to "stop" channel. For lifetime measurement the "start" channel of the correlator was connected to the sync trigger of the pulsed laser and "stop" channel was connected to the SSPD.

## $g^{(2)}(\tau)$ autocorrelation function measurements and background compensation

In the experiment actual signal–to–noise ratio is rather low. Due to the metamaterial or other substrate fluorescence the measured level of $g^{(2)}(\tau)$ may be relatively high even if we deal with a single-photon emitter. In order to compensate for the background noise, we measured background level next to an NV-center and then recalculated $g^{(2)}(\tau)$. By definition,

$$g^{(2)}(\tau) = \frac{\langle S_1(t)S_2(t+\tau)\rangle}{\langle S_1(t)\rangle\langle S_2(t+\tau)\rangle}, \qquad (2)$$

where $S_i(t)$ is the light intensity at each detector. In the presence of noise $N(t)$, total signal on the detector would be $I(t) = S(t) + N(t)$ and measured $\hat{g}^{(2)}(\tau)$, assuming that the noise is statistically independent of the signal, will be:

$$\hat{g}^{(2)}(\tau) = \frac{\langle (S_1(t) + N_1(t))(S_2(t+\tau) + N_2(t+\tau))\rangle}{\langle S_1(t) + N_1(t)\rangle\langle S_2(t+\tau) + N_2(t+\tau)\rangle} = \qquad (3)$$

$$= g^{(2)}(\tau)\frac{\langle S_1(t)\rangle\langle S_2(t+\tau)\rangle + \langle N_1(t)\rangle\langle N_2(t+\tau)\rangle + \langle N_1(t)\rangle\langle S_2(t+\tau)\rangle + \langle N_2(t+\tau)\rangle\langle S_2(t+\tau)\rangle}{\langle S_1(t)\rangle\langle S_2(t+\tau)\rangle + \langle N_1(t)\rangle\langle S_2(t+\tau)\rangle + \langle N_2(t+\tau)\rangle\langle S_2(t+\tau)\rangle + \langle N_1(t)\rangle\langle N_2(t+\tau)\rangle}.$$

This allows to find true $g^{(2)}(\tau)$ if the average values $\langle I_i(t)\rangle$, $\langle N_i(t)\rangle$ were measured while taking correlation data:

$$g^{(2)}(\tau) = \hat{g}^{(2)}(\tau)\Big(\frac{\langle S_1\rangle\langle S_2\rangle + \langle N_1\rangle\langle S_2\rangle + \langle N_2\rangle\langle S_2\rangle + \langle N_1\rangle\langle N_2\rangle}{\langle S_1\rangle\langle S_2\rangle} -$$

$$- \frac{\langle N_1\rangle\langle N_2\rangle + \langle N_1\rangle\langle S_2\rangle + \langle N_2\rangle\langle S_2\rangle}{\langle S_1\rangle\langle S_2\rangle}\Big), \langle S_i\rangle = \langle I_i\rangle - \langle N_i\rangle, \ i=1,2. \qquad (4)$$

This procedure was used in all our measurements of correlation function. Uncompensated data is presented in Fig. 4. Unfortunately due to the fact that background is measured at different location, complete compensation is not possible. Normally NV nanodiamond serves as a scattering center, therefore increase in local level of background leads to uncompensated background. In the main text, model for $g^{(2)}(\tau)$ in Fig. 2 was adjusted for the residual background level, so the effective signal–to–noise ratio is around 100. Actual single to noise ratio in the experiment is about 2.3. Modeling of $g^{(2)}(\tau)$ for uncompensated data is shown in the Fig. 6.

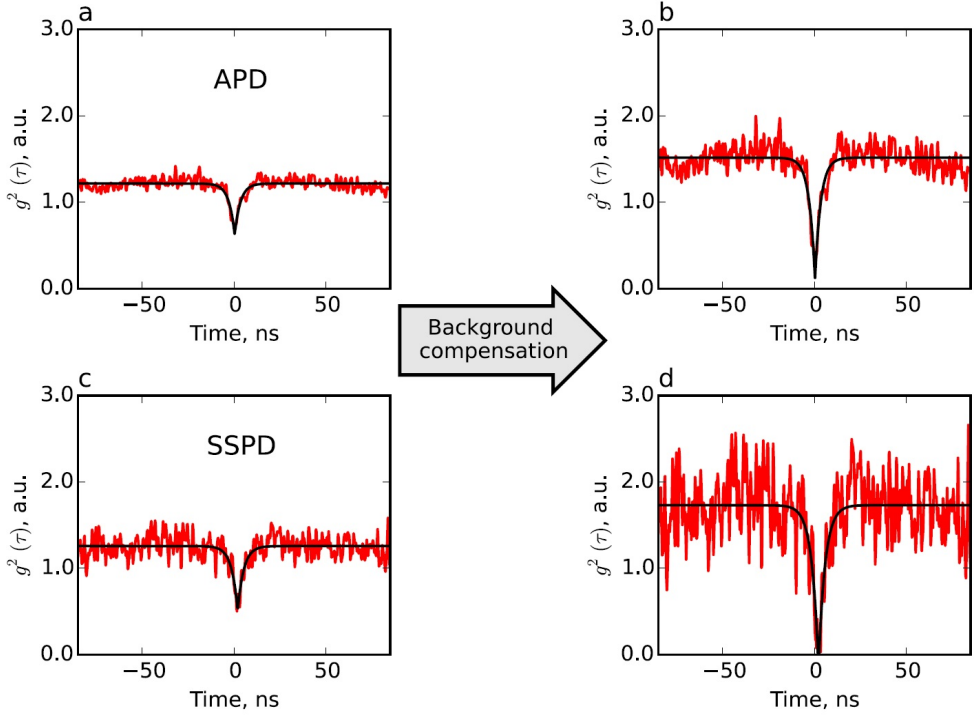

Fig. 4. Measured g$^{(2)}(\tau)$ from a single NV center on top of hyperbolic metamaterial with APD (top) and SSPD (bottom). Left part of the figure represents g$^{(2)}(\tau)$ "as measured", right after background compensation procedure. Effect on g$^{(2)}(\tau)$ level by signal–to–noise ratio of the single photon emitter with respect to background correspond to effective improvement from s/n=3 to s/n=100

## Statistical modeling of g$^{(2)}(\tau)$ autocorrelation function measurements

Incoherent pumping model (Fig. 5) with pumping rate $R$ and spontaneous decay rate $\gamma$ will give the following rate equation for a population of two active levels:

$$\frac{dn_e}{dt} = R n_g - \gamma n_e \tag{5}$$

$$n_e + n_g = 1, \tag{6}$$

where $n_e$ and $n_g$ are the populations of level $|e\rangle$ and $|g\rangle$, respectively. Here we assume, that level $|3\rangle$ in Fig. 5 is not populated and relaxes to level $|e\rangle$ within negligibly short time. The solution of this equation, assuming that a photon was just emitted ($n_e(t=0) = 0$), gives:

$$n(t) = \frac{r}{r+\gamma}(1 - \exp(-t(r+\gamma))) \tag{7}$$

$$n(t) = \frac{\tau_{exc}^{-1}}{\tau_{exc}^{-1} + \tau_{decay}^{-1}}\left(1 - \exp\left(-t\left(\tau_{exc}^{-1} + \tau_{decay}^{-1}\right)\right)\right). \tag{8}$$

This dependence in fact represents a probability of the photon next to the one detected to be emitted and, therefore, represents the form of the g$^{(2)}(\tau)$ correlation function. The normalization was done using standard definition given by Eq. (2). Here we do not consider ionization process of NV center due to presence of a dark state, which leads to increased level of g$^{(2)}$($\tau$) since this process do not affect value of g$^{(2)}$(0). The background level and detector noise was modeled using Eq. (4). The jitter was included assuming Gaussian distribution:

$$P(t) = \frac{1}{\sqrt{2\Pi}\sigma} \exp\left(-\frac{1}{2}\left(\frac{t}{\sigma}\right)^2\right) \quad (9)$$

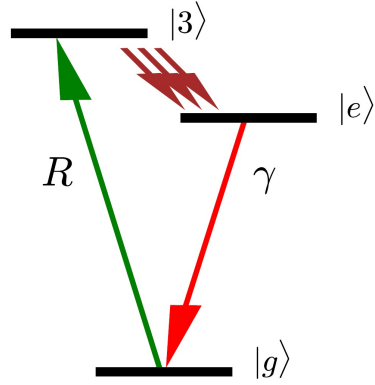

Fig. 5. Simplified level diagram for incoherent pumping model.

By convolving expression for g$^{(2)}$($\tau$) from Eq. (4) with Eq. (9):

$$g^{(2)}(t) = \int_{-\infty}^{+\infty} g^{(2)}(\tau + t) P(t) \, dt \quad (10)$$

The g$^{(2)}$(0) was used to plot dependences in Fig. 2(d) of the main text and Fig. 6.

**Dynamic Range**

Detector step behavior model: under continuous-wave illumination at high count rates, detector's output pulse will mostly fire before complete detector relaxation [61]. The latter and the fact that detector's output is AC-coupled requires accurate discriminator level adjustment for each point. Digital oscilloscope was used to count pulses in the following way: waveforms were saved on a USB stick and analyzed in python program with visual discriminator level control. Fig. 3(c) represents pulse waveform points, obtained from the oscilloscope. Pulses are aligned by positive edge. Axis Y corresponds to voltage on the 50 Ohm resistor shunting the detector 200 times amplified. Characteristic pulse form is fitted by function $U(t) = U + U_0 \exp(-t)$, where $U_0$ voltage on the detector. Obtained values of $U_0$ are represented in Fig. 3(d) along with a count rate. We made a computer model of the detector using detector's efficiency $\nu(I)$ and detector's current pulse decay time $\tau$. We calculated an output pulse frequency for $N_{in}$ - incoming photon flux and $\bar{I}_0$ - mean detector's current, maintained by power supply circuit.

Definitions: $N_{in}$ - input photon flux; $N_{out}$ - output pulse frequency; $\nu(I)$ - detector quantum efficiency; $\bar{I}$ - mean detector current; $I_w$ - detector operation current, that flows through detector without detector firing; $\tau$ - characteristic pulse decay time.

$I(I_w, t) = I_w (1 - \exp(-t/\tau))$ - detector's current as a function of time just after firing.

$P_{reg}(t, I_w)$ - firing probability since $t$ after previous firing at operating current $I_w$.

$p_{reg}(t, I_w)$ - probability density since $t$ after previous firing at operating current $I_w$.

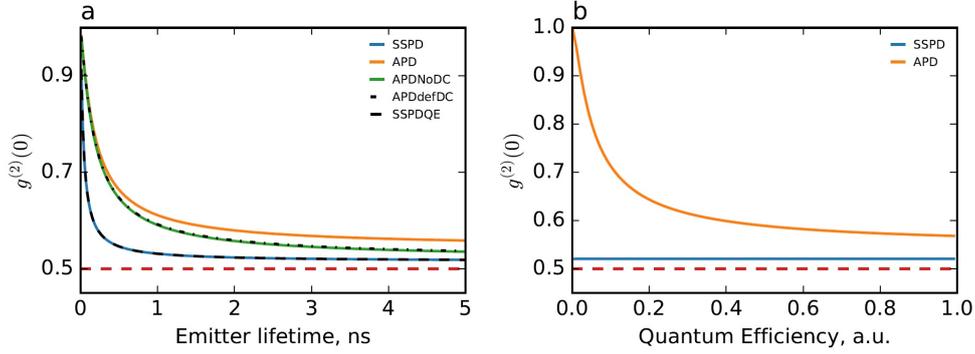

Fig. 6. Results of $g^{(2)}(0)$ calculations based on the model described for realistic experimental parameters. Here signal coming to a detector was assumed to be 80k counts per second, noise level was set to 37k counts per second, giving signal-to-noise ratio of 2.3. a) $g^{(2)}(0)$ versus emitter lifetime. Orange curve represents an estimate for real APD with measured dark counts rate, green corresponds to ideal APD with no dark counts but same jitter of 350 ps as real one, black dash doted line represents APD with dark counts specified in data sheet; black and blue correspond to the SSPD with and without dark counts. b) level of $g^{(2)}(0)$ for a given lifetime of 3 ns versus quantum efficiency of the detector. SSPD was only calculated starting from 0.001 quantum efficiency which is still enough to overcome dark counts with a signal assumed above.

The probability of the event that detector fires within the range $(t, t+dt)$ is equal to a product of the probability of not firing before time t and probability of firing within the range:

$$p_{reg}(t, I_w, N_{in}) \, dt = \left(1 - P_{reg}(t, I_w)\right) \nu I(I_w, t) N_{in} dt \tag{11}$$

Taking into account that

$$p_{reg} = \frac{dP_{reg}}{dt} \tag{12}$$

we obtain a numerically solved equation for $p_{reg}$.

Mean current for periodically pulses with period $T$ :

$$I_{avg}(T, I_w) = \frac{1}{T} \int_0^T I(I_w, t) \, dt = I_w \left(1 + \frac{t}{\tau} \left(\exp\left(-\frac{t}{\tau}\right) - 1\right)\right) \tag{13}$$

Mean current for pulses with time between them distributed randomly with probability density $p_{reg}$ turns out to be:

$$\bar{I}(I_w) = \frac{\int_0^\infty I_{avg}(t, I_w) p_{reg}(t, I_w) \, dt}{\int_0^\infty t p_{reg}(t, I_w) \, dt} \tag{14}$$

For operating current we have equation: $\bar{I}_0 = \bar{I}(I_w)$

Count rate is inverted mean time between pulses:

$$N_{out} = \frac{1}{\int_0^\infty t p_{reg}(t, I_w) \, dt} \tag{15}$$

## Funding

This work was partially supported by AFOSR grant (FA9550–12–1–0024), NSF–MRSEC grant (DMR–1120923), Russian Foundation for Basic Research grant 14–29–07127, scientific state tasks No. 960, 3.2655.2014/K, 3.1846.2014/K and grant by the Ministry of Education and Science of the Russian Federation No. 14.B25.31.0007.

**Acknowledgments**

The authors would like to thank Nathaniel Kinsey and Samuel Peana for their kind assistance with manuscript preparation, Roman Ozhegov for explanation of SSPD DC bias unit operation and Yuliya Korneeva for useful discussion of optical cavity performance.